\def\be{\begin{equation}}
\def\ee{\end{equation}}
\def\bea{\begin{eqnarray}}
\def\eea{\end{eqnarray}}
\def\bse{\begin{subequations}}
\def\ese{\end{subequations}}

\def\jc{j_{\,\text{c}}}
\def\Hc{H_{\text{c}}}
\def\Tc{T_{\text{c}}}
\def\mun{\mu_{\text{n}}}
\def\chin{\chi_{\text{n}}}
\def\chis{\chi_{\text{s}}}
\def\mus{\mu_{\text{s}}}
\documentclass[prb,twocolumn,showpacs,amsmath,amssymb,eqsecnum,preprintnumbers]{revtex4}
\usepackage{graphicx}
\usepackage{dcolumn}
\usepackage{bm}
\begin{document}
\title{Nearly Ferromagnetic Superconductors}
\author{Qi Li and D. Belitz}
\affiliation{Department of Physics and Materials Science Institute, University
                    of Oregon, Eugene, OR 97403, USA}
\author{T.R. Kirkpatrick}
\affiliation{Institute for Physical Science and Technology,
                    and Department of Physics, University of Maryland, College Park,
                    MD 20742, USA}
\date{\today}

\begin{abstract}
The electromagnetic properties of superconductors near a ferromagnetic
instability are investigated by means of a generalized Ginzburg-Landau theory.
It is found that the magnetic flux expulsion capability of the superconductor
gets stronger, in a well-defined sense, as the normal-state magnetic
susceptibility increases.
The temperature dependencies of the London penetration depth, the critical
fields, and the critical current are all strongly affected by ferromagnetic
fluctuations. In particular, for the critical current we find a temperature
exponent $\alpha \approx 2$ over an appreciable temperature range. The extent
to which proximity to magnetic criticality may be a viable explanation for
recent observations in MgCNi microfibers, which find $\alpha \approx 2$, is
discussed.
\end{abstract}


\maketitle

\section{Introduction}
\label{sec:I}

The coexistence of ferromagnetism and superconductivity has received
substantial attention over the years. Around 1980, such states were
predicted\cite{Blount_Varma_1979, Greenside_Blount_Varma_1981,
Kuper_Revzen_Ron_1980} and observed,\cite{Moncton_et_al_1980, Lynn_et_al_1981}
and the topic later received renewed interest in the context of experimental
observations in rare earth borocarbides.\cite{Ng_Varma_1997} More recently,
interest in this subject has been revived by the observation of coexisting
superconductivity and ferromagnetism in UGe$_2$\cite{Huxley_et_al_2001,
Saxena_et_al_2000} and URhGe,\cite{Aoki_et_al_2001} where both types of order
are believed to be due to electrons in the same band. Recent theoretical
attention has centered on the structure of the phase
diagram,\cite{Kirkpatrick_Belitz_2003a} on the existence of spontaneous flux
lattices,\cite{Ng_Varma_1997, Radzihovsky_et_al_2001, Tewari_et_al_2004} and on
the question of spin-triplet versus spin-singlet
superconductivity.\cite{Belitz_Kirkpatrick_2004}

In contrast, less is known about the properties of superconductors on the
paramagnetic side of, but close to, a ferromagnetic instability. We will refer
to ``paramagnetic superconductors'' to describe systems in this regime,
although the superconductivity of course leads to the usual strong diamagnetic
effects. Such paramagnetic superconductors include systems below the
superconducting transition temperature, but above the temperature below which
coexistence of superconductivity and ferromagnetism occurs, as well as systems
that never develop ferromagnetism, but are close to a ferromagnetic instability
in some direction in parameter space other than temperature. An example of the
latter is believed to be the non-oxide perovskite MgCNi$_3$, which
superconducts below a critical temperature $T_{\text{c}}\approx
8\,{\text{K}}$.\cite{He_et_al_2001} There is no evidence for a ferromagnetic
phase in this material, but it has been suggested that a ferromagnetic ground
state can be reached upon a relatively small amount of hole
doping.\cite{Rosner_et_al_2002} This system may thus be close to a
ferromagnetic instability everywhere in its superconducting phase.

A recent study of MgCNi$_3$ microfibers, with $T_{\text{c}} = 7.8\,{\text{K}}$,
has revealed an anomalous temperature dependence of the critical current
density $\jc$.\cite{Young_Moldovan_Adams_2004} The critical current density
vanishes at $T_{\text{c}}$ according to a power law $\jc \propto \vert T -
T_{\text{c}}\vert^{\alpha}$, with $\alpha=2$ between about 1\% and 10\% away
from the critical point, and no crossover to the usual Ginzburg-Landau
behavior, which predicts $\alpha=3/2$. The authors of Ref.
\onlinecite{Young_Moldovan_Adams_2004} have ruled out morphological effects as
an explanation, which raises the question whether proximity to a ferromagnetic
state may be responsible. Indeed, since ferromagnetic fluctuations are expected
to weaken (singlet) superconductivity, this is a plausible suggestion for the
origin of the weaker-than-expected temperature dependence of $\jc$.

The probable proximity to ferromagnetism has led to a debate about the nature
and symmetry of the pairing in MgCNi$_3$.\cite{Prozorov_Gianetta_2006} This
point has not been settled; some experimental evidence points to conventional
s-wave pairing; other, to a superconducting order parameter with nodes. The
nature of the pairing in the other materials mentioned above has not been
unambiguously determined either. In this paper we will focus on the behavior
close to $T_{\text{c}}$, which is qualitatively independent of the symmetry of
the order parameter and thus expected to be the same for all nearly
ferromagnetic superconductors. We use a generalized Ginzburg-Landau theory to
theoretically investigate the electrodynamic properties of a superconductor as
a ferromagnetic instability is approached. We treat the superconductivity in
the usual mean-field approximation, but the magnetic critical behavior exactly
in a scaling sense. Somewhat counter-intuitively, strong magnetic fluctuations
make, in a well-defined sense, the superconductivity more robust in certain
respects. In particular, the penetration depth becomes anomalously short.
The thermodynamic critical field, on the other hand, becomes weaker, as one
might intuitively expect. The temperature dependencies of the critical field
$\Hc$ and the penetration depth $\lambda$ depend on the magnetic critical
exponents $\delta$ and $\gamma$, respectively. For the critical current $\jc
\propto \Hc/\lambda$, this results in an exponent $\alpha$ between 1.5 (the
Ginzburg-Landau result) and 2.16 in various temperature regimes. We will
discuss both the existing experimental observations, and predictions for new
experiments, in the light of these results.

This paper is organized as follows. In Sec.\ \ref{sec:II} we give elementary
phenomenological arguments for the dependence of the thermodynamic critical
field, the penetration depth, and the critical current density, on a constant
normal-state magnetic permeability $\mun$. We then generalize these results to
the magnetically critical case, where one needs to distinguish between $\mun$
and the spin susceptibility $\mus$ in a superconduting state, and both $\mun$
and $\mus$ become nonanalytic functions of various control parameters. In Sec.\
\ref{sec:III} we derive these results from a generalized Ginzburg-Landau
theory, and in Sec.\ \ref{sec:IV} we give a discussion of our results.

\section{Phenomenological Arguments}
\label{sec:II}

\subsection{Paramagnetic systems}
\label{subsec:II.A}

We are interested in the electromagnetic properties of superconductors with
ferromagnetic fluctuations. We denote the normal-state spin susceptibility,
which describes the response of the spin degrees of freedom to an external
magnetic field in the absence of superconductivity, by $\chin$, and the
corresponding spin permeability by $\mun = 1 + 4\pi\chin$. This is in contrast
to the spin permeability $\mus = 1 + 4\pi\chis$, which includes the effects of
the superconductivity on the spin response, and the magnetic permeability $\mu
= 1 + 4\pi\chi$, which describes the response of the total magnetization,
including the diamagnetic part. It is instructive to first recall the
dependence of superconducting properties on a constant $\mun\neq 1$, neglecting
the distinction between $\mun$ and $\mus$.\cite{Ginzburg_1956,
Chen_Lubensky_Nelson_1978} This can be done by means of elementary arguments.

\subsubsection{Thermodynamic critical field}
\label{subsubsec:II.A.1}

Consider the free energy density $f$ of a system in a magnetic field. It obeys
\be
df = df(H=0) + \frac{1}{4\pi}\,H\,dB,
\label{eq:2.1}
\ee
where $H$ is the thermodynamic magnetic field, and $B$ is the magnetic
induction. For the sake of simplicity, we ignore the vector nature of various
quantities in our free energy considerations. For fixed $B$, $f$ is the
appropriate thermodynamic potential whose minimum determines the equilibrium
state. However, in an experiment $H$ is fixed, since
$(c/4\pi){\bm\nabla}\times{\bm H} = {\bm j}_{\,\text{ext}}$ is the external
current density, and only the latter is experimentally controlled. One
therefore must perform a Legendre transform to a thermodynamic potential $g = f
-BH/4\pi$,\cite{Tinkham_1975, DeGennes_1989} which obeys
\be
dg = df(H=0) -\frac{1}{4\pi}\,B\,dH.
\label{eq:2.2}
\ee
In a paramagnetic phase, including paramagnetic superconductors, the relation
between $B$ and $H$ is
\be
B = H + 4\pi M = (1 + 4\pi\chi)H = \mu H,
\label{eq:2.3}
\ee
with $M$ the magnetization, $\chi(T,H)$ the magnetic susceptibility, and $\mu =
1 + 4\pi\chi$ the magnetic permeability. Integration of Eq.\ (\ref{eq:2.2})
yields
\be
g(T,H) = f(T,H=0) - \frac{1}{4\pi}\int_0^H dh\,\left[1 + 4\pi\chi(T,h)\right]h.
\label{eq:2.4}
\ee
This is generally valid. In a superconducting Meissner state, $B=0$, and hence
$\chi=-1/4\pi$ (ideal diamagnetism), and $f(T,H=0) = f_0 + t\vert\psi\vert^2/2
+ u\vert\psi\vert^4/4$, with $f_0$ the free energy density of the normal state,
$\psi$ the superconducting order parameter, $t \propto (T - \Tc)/\Tc$ the
dimensionless distance from the superconducting critical point, and $u$ a
parameter. In a normal metal far from a magnetic instability, and ignoring
normal-state diamagnetic effects, $\chi(T,H) \approx \text{const.} \equiv
\chin$, or $\mun = 1 + 4\pi\chin = \text{const.}$, and $f(T,H=0) = f_0$. In a
normal metal close to a ferromagnetic critical point, $\chi$ is a complicated
function of $T$ and $H$.

Now consider a superconductor with $\mun = \text{const}$. According to Eq.\
(\ref{eq:2.4}), the magnetic energy density gained by the system allowing
magnetic flux to penetrate, i.e., the free energy density difference between
the Meissner state with $B=0$ and the normal state with $B=\mun H$, is
$E_{\text{m}}/V = \mun H^2/8\pi$. By contrast, the condensation energy density
gained by the system becoming a superconductor is $E_{\text{cond}}/V = t^2/4u$.
The thermodynamic critical field, which is defined by these two energies being
equal, is thus
\be
\Hc = \sqrt{2\pi/u}\ \vert t\vert/\sqrt{\mun} = \Hc^0/\sqrt{\mun},
\label{eq:2.5}
\ee
with $\Hc^0 = \sqrt{2\pi/u}\,\vert t\vert$ the critical field for a system with
$\mu_{\text{n}}=1$. An increase in $\mun$ thus decreases the critical field, as
one might expect since the externally applied field is amplified inside the
material.

\subsubsection{London penetration depth}
\label{subsubsec:II.A.2}

\begin{figure}[t,h]
\vskip -0mm
\includegraphics[width=8.0cm]{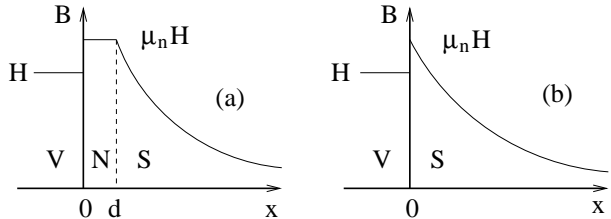}
\caption{Magnetic induction schematically as a function of position at a vacuum
(V) - normal metal (N) - superconductor interface (S) (a), and at a vacuum -
superconductor interface (b).}
\label{fig:1}
\end{figure}

The dependence of the London penetration depth $\lambda$ on $\mu_{\text{n}}$ is
intuitively less obvious. Consider a large superconducting sample, with linear
dimension $L$, surrounded by vacuum and subject to a homogeneous external
magnetic field ${\bm H} = (0,0,H)$ in $z$-direction. Along the left edge of the
sample, the magnetic induction will be of the form ${\bm B}({\bm x}) =
(0,0,B(x))$ with $B(x) = B_0\,e^{-x/\lambda}$ ($x>0$). To determine $B_0$,
imagine a thin (thickness $d$) layer of normal conducting material around the
superconductor. Except for the superconductivity, the normal layer should have
the same properties as the superconductor, in particular, a magnetic
permeability $\mu_{\text{n}}$. Then we have $B=H$ in vacuum, and $B =
\mu_{\text{n}}H$ inside the normal layer, see Fig.\ \ref{fig:1}(a). Now let
$d\to 0$. Then we have (Fig.\ \ref{fig:1}(b))
\be
B(x) = \begin{cases} H \quad \text{for} \quad x<0 \cr
                     \mu_{\text{n}}H\,e^{-x/\lambda} \quad \text{for} \quad x\ge 0
       \end{cases}\ .
\label{eq:2.6}
\ee

Now consider the current density associated with $B(x)$. From Ampere's law we
have
\bse
\label{eqs:2.7}
\be
{\bm j}({\bm x}) = \frac{c}{4\pi}{\bm\nabla}\times{\bm B}({\bm x}) =
(0,j(x),0),
\label{eq:2.7a}
\ee
with $c$ the speed of light and
\be
j(x) = \frac{c}{4\pi\lambda}\,B(x).
\label{eq:2.7b}
\ee
\ese
This is the {\em total} current density. It has three contributions, namely,
the supercurrent density ${\bm j}_{\,\text{sc}}$, the spin or magnetization
current ${\bm j}_{\,\text{spin}} = c{\bm\nabla}\times{\bm M}$, with ${\bm M}$
the spin contribution to the total magnetization, and the external current
density ${\bm j}_{\,\text{ext}} = c{\bm\nabla}\times{\bm H}/4\pi$. The latter
vanishes in the case we are considering. In a normal metal, the spin current is
the only contribution if we ignore normal-state diamagnetic effects. The spin
or normal-state susceptibility $\chi_{\text{n}}$ is defined as the response of
${\bm M}$ to the total magnetic induction ${\bm B}$ minus the contribution to
${\bm B}$ of ${\bm M}$ itself,
\be
{\bm M} = \chi_{\text{n}}({\bm B} - 4\pi{\bm M}) = (\chin/\mun)\,{\bm B}.
\label{eq:2.8}
\ee
For the supercurrent density ${\bm j}_{\,\text{sc}} = {\bm j} - {\bm
j}_{\,\text{spin}}$ this implies
\bse
\label{eqs:2.9}
\be
{\bm j}_{\,\text{sc}} = \frac{c}{4\pi\mu_{\text{n}}}\,{\bm\nabla}\times{\bm
B}({\bm x}) = (0,j_{\,\text{sc}}(x),0),
\label{eq:2.9a}
\ee
with
\be
j_{\,\text{sc}}(x) = \frac{c}{4\pi\mu_{\text{n}}\lambda}\,B(x).
\label{eq:2.9b}
\ee
\ese
Now consider one surface (area $L^2$) of the sample. Neglecting corner effects,
and for $\lambda \ll L$, the total magnetic flux $\Phi$ through that surface is
\be
\Phi = \int_0^L dy \int_0^L dx\,B(x) \approx L\int_0^{\infty}dx\,B(x) =
L\lambda\mu_{\text{n}}H.
\label{eq:2.10}
\ee
On the other hand, the total supercurrent flowing near that surface is, from
Eq.\ (\ref{eq:2.9b}),
\be
I_{\text{sc}} = \int d{\bm x}\,j_{\,\text{sc}} \approx
L^2\int_0^{\infty}dx\,j_{\,\text{sc}}(x) =
\frac{c}{4\pi\mu_{\text{n}}}\,\frac{L}{\lambda}\,\Phi.
\label{eq:2.11}
\ee
We thus can write the flux
\be
\Phi = \frac{4\pi\mu_{\text{n}}}{c}\,\frac{\lambda}{L}\,I_{\text{sc}} =
\frac{4\pi\mu_{\text{n}}}{c}\,\frac{\lambda}{L}\,Nqv,
\label{eq:2.12}
\ee
where $N$ is the number of supercurrent carrying particles, $q$ is their
charge, and $v$ is their velocity. If $m$ is their mass, then $E_{\text{kin}} =
Nmv^2/2$ is the total kinetic energy of the supercurrent. The flux can thus be
written
\be
\Phi =
\frac{4\pi\mu_{\text{n}}}{c}\,\frac{\lambda}{L}\,q\sqrt{2/m}\sqrt{N}\sqrt{E_{\text{kin}}}.
\label{eq:2.13}
\ee
Now we make two observations. First, $N\approx L^2\lambda n$, with $n =
\vert\psi\vert^2$ the particle number density. Second, at the critical field
strength the kinetic energy of the supercurrent must equal the condensation
energy in the region where the current is flowing, which is (see Sec.\
\ref{subsec:II.A}) $E_{\text{cond}} = L^2\lambda t^2/u$. With $\lambda_0$ the
London penetration depth for $\mu_{\text{n}}=1$,
\be
\lambda_0 = \sqrt{mc^2/4\pi q^2\vert\psi\vert^2} \propto 1/\vert t\vert^{1/2},
\label{eq:2.14}
\ee
this allows to write the flux at the critical field
\be
\Phi_{\text{c}} = \lambda L H_{\text{c}}^0\mu_{\text{n}}\lambda/\lambda_0 =
L\lambda\, \mu_{\text{n}}\Hc,
\label{eq:2.15}
\ee
where the first equality follows from Eq.\ (\ref{eq:2.13}), and the second one
from Eq.\ (\ref{eq:2.10}). We thus obtain
\bse
\label{eqs:2.16}
\be
H_{\text{c}}/\lambda = H_{\text{c}}^0/\lambda_0\ ,
\label{eq:2.16a}
\ee
or\cite{Ginzburg_1956, Chen_Lubensky_Nelson_1978}
\be
\lambda = \lambda_0/\sqrt{\mu_{\text{n}}}\ .
\label{eq:2.16b}
\ee
\ese
The penetration depth thus {\em decreases} with increasing $\mu_{\text{n}}$, as
does the critical field. This is somewhat counterintuitive, as it implies that
the superconductivity becomes in some sense more robust. It also implies that a
large normal-state magnetic permeability will make the superconductor
necessarily of type I.\cite{Greenside_Blount_Varma_1981} We will come back to
this observation.

Notice that the above derivation relies only on very general energetic
considerations and on Ampere's law. Also notice that it uses an identity at the
critical field strength, where the superconductivity vanishes. This is fine for
$\Hc$, but the penetration depth is a property of the superconducting state,
and hence the use of $\mun$ is not quite appropriate for this quantity, except
in the limit $\lambda \to \infty$. More generally, $\lambda$ depends on $\mus$,
which in turn depends on the superconducting properties. This makes no
difference deep inside the paramagnetic superconducting phase, and Eq.\
(\ref{eq:2.16b}) is valid there. However, as we will see it makes a crucial
difference close to a ferromagnetic instability.

\subsubsection{Critical current}
\label{subsubsec:II.A.3}

In order to discuss the critical current, we assume a thin-wire geometry with
wire radius $R$.\cite{Tinkham_1975, DeGennes_1989} The supercurrent density,
which is the total current density minus the spin current density, can be
written as a generalization of Eq.\ (\ref{eq:2.9a}),
\bea
j_{\text{sc}}({\bm x}) &=& \frac{c}{4\pi}\,\left[{\bm\nabla}\times{\bm B}({\bm
x}) - 4\pi{\bm\nabla}\times{\bm M}({\bm x})\right]
\nonumber\\
&=& \frac{c}{4\pi\mu({\bm x})}\,{\bm\nabla}\times{\bm B}({\bm x}),
\label{eq:2.17}
\eea
where we have used Eq.\ (\ref{eq:2.3}) and $\mu({\bm x})$ is the local magnetic
susceptibility. Now integrate over the cross section of the wire. Assuming a
homogeneous current density within a distance $\lambda$ from the surface, and
using Gauss's theorem on the right-hand side, we have
\[
2\pi R\lambda j_{\text{sc}} = \frac{c}{4\pi\mun} \oint
d{\bm\ell}\cdot\left({\bm\nabla}\times{\bm B}({\bm x})\right) = \frac{c}{2}\,RH
\]
where we have used Eq.\ (\ref{eq:2.6}). The
critical current density $\jc$ is the one that produces the thermodynamic
critical field $\Hc$, which yields the familiar London theory
result\cite{Tinkham_1975}
\be
\jc = c H_{\text{c}}/4\pi\lambda.
\label{eq:2.18}
\ee
This result is plausible: Dimensionally, $\jc$ must be a magnetic field divided
by a length. The relevant length scale is the thickness of the area that
supports diamagnetic currents, which is $\lambda$. The relevant field scale
should be the field that corresponds to the condensation energy, which is
$H_{\text{c}}$. To the extent that $\mus\approx\mun = \text{const.}$, as we
have assumed in Sec.\ \ref{subsubsec:II.A.2}, Eq.\ (\ref{eq:2.16a}) implies
that $\jc$ is independent of $\mun$,
\be
\jc = \jc^{\,0}.
\label{eq:2.19}
\ee
As we will see below, this result changes drastically in the vicinity of a
ferromagnetic instability.

\subsection{Systems at a ferromagnetic instability}
\label{subsec:II.B}

In the vicinity of a ferromagnetic instability of the normal metal, the normal
state magnetic susceptibility $\chin$, and hence the permeability $\mun$,
become large and diverge as the phase transition is approached. At a
ferromagnetic critical point, the region of linear response shrinks to zero,
and $\chin$ and $\mun$ become strongly field dependent. This field dependence
is characterized by the critical exponent $\delta$,\cite{Ma_1976}
\be
\mun \approx \chin \propto H^{1/\delta -1}.
\label{eq:2.20}
\ee
The value of $\delta$ depends on the universality class the particular magnetic
system belongs to. For all realistic universality classes, $\delta \approx 5$,
whereas in Landau or mean-field theory, $\delta=3$.\cite{Zinn-Justin_1996}
Substituting Eq.\ (\ref{eq:2.20}) into Eq.\ (\ref{eq:2.5}), we find for the
thermodynamic critical field
\be
\Hc \propto \vert t\vert^{2\delta/(\delta+1)}
\label{eq:2.21}
\ee
This result holds for a system where the distance $t$ from the superconducting
critical point can be changed while the system remains tuned to magnetic
criticality (more precisely, to the parameter values where magnetic criticality
would occur in the absence of superconductivity). Generically, the
dimensionless distance $r$ from magnetic criticality will change as well if $t$
is changed, and we will discuss such more realistic situations in Sec.\
\ref{sec:IV}.

For the penetration depth, the situation is more complicated. In contrast to
$\Hc$, which compares the normal-state magnetic energy with the superconducting
condensation energy that has nothing to do with spin magnetism, $\lambda$ is
entirely a property of the superconducting state, and the feedback of the
superconductivity on the spin susceptibility, or the difference between $\mun$
and $\mus$, cannot be neglected. As a result of this feedback, the magnetic
transition in the presence of superconductivity does not occur at $r=0$, but
rather at a value $r \propto
-\xi_{\text{m}}^0/\lambda_0$.\cite{Blount_Varma_1979} Here $\xi_{\text{m}}^0$
is the magnetic correlation length at zero temperature. This suggests that the
spin susceptibility at $r=0$ will be effectively $\chis \propto
\lambda_0/\xi_{\text{m}}^0 \gg 1$ in a mean-field approximation. More
generally, one has $\mus \approx \chis \propto
(\lambda_0/\xi_{\text{m}}^0)^\gamma$, with $\gamma$ another critical exponent.
Using this in Eq.\ (\ref{eq:2.16b}) with $\mun$ replaced by $\mus$, we obtain
\be
\lambda \propto
\lambda_0^{1-\,\gamma/2}\left(\xi_{\text{m}}^0\right)^{\gamma/2} \propto \vert
t\vert^{-1/2 + \gamma/4}.
\label{eq:2.22}
\ee
Since $\gamma \approx 1.4 > 0$ for ferromagnetic
systems,\cite{Zinn-Justin_1996} this implies that the penetration depth at
magnetic criticality is anomalously short. Close to the superconducting
transition, the superconductor will therefore also be of type I, in agreement
with a conclusion drawn from studying the ferromagnetic
phase.\cite{Greenside_Blount_Varma_1981}

For the critical current density, Eqs.\ (\ref{eq:2.18}), (\ref{eq:2.21}), and
(\ref{eq:2.22}) predict
\bse
\label{eqs:2.23}
\be
\jc \propto \vert t\vert^{\alpha}
\label{eq:2.23a}
\ee
with
\be
\alpha = 2\delta/(\delta + 1) + 1/2 - \gamma/4 \quad,\quad (r=0) \quad.
\label{eq:2.23b}
\ee
With $\delta \approx 5$ and $\gamma \approx 1.4$ this yields $\alpha \approx
1.8$, in contrast to the Ginzburg-Landau result $\alpha = 3/2$.

These results hold at $r=0$, and again we have assumed that $t$ can be varied
independently of $r$. Let us relax the former condition. From the above
argument for the effective value of $\chis$ at $r=0$ it also follows that Eq.\
(\ref{eq:2.22}) is valid only for $\vert r\vert < \xi_{\text{m}}^0/\lambda_0$.
Since $\xi_{\text{m}}^0$ is typically on the order of a few $\AA$, while
$\lambda_0$ is typically several hundred $\AA$ or even larger even at zero
temperature, and diverges as $\vert t\vert^{-1/2}$ for $t\to 0$, this is a very
small range. By contrast, Eq.\ (\ref{eq:2.20}) can be valid for $r$ as large as
several percent, provided $H$ is not too small. Not too close to
$T_{\text{c}}$, where $\Hc$ goes to zero, Eq.\ (\ref{eq:2.21}) can thus be
valid in a substantial $r$-range, while $\lambda = \lambda_0/\sqrt{\mun}$
except in an extremely small interval around $r=0$. In that case,
\be
\alpha = 2\delta/(\delta + 1) + 1/2\quad,\quad (1\gg r \gg
\xi_{\text{m}}^0/\lambda_0)\quad,
\label{eq:2.23c}
\ee
\ese
which yields $\alpha \approx 2.17$ if $\delta \approx 5$.

Finally, at larger values of $r$, or sufficiently close to $T_{\text{c}}$ that
$\Hc$ is small enough to invalidate Eq.\ (\ref{eq:2.20}), we are back to the
paramagnetic case, Eq.\ (\ref{eq:2.19}) holds, and thus $\alpha = 3/2$.

One thus faces a rather complicated situation, where the exponent $\alpha$ can
take on values between the Ginzburg-Landau value 3/2 and a value larger than
$2$, Eq.\ (\ref{eq:2.23c}), depending on various parameters that are not easy
to control or even determine experimentally. We will discuss this in more
detail in Sec.\ \ref{sec:IV}. Before we do so, in the following section we will
give a more technical and more detailed derivation of all of our results.

\section{Generalized Ginzburg-Landau Theory}
\label{sec:III}

We now consider a coupled field theory that describes both superconducting and
spin degrees of freedom in order to derive the above results from a more
microscopic level and gain a deeper understanding of their origin.
Specifically, we consider a generalization of the usual Ginzburg-Landau
equations that includes the spin degrees of freedom. Far from magnetic
criticality, the latter can be integrated out to yield ordinary Ginzburg-Landau
theory with $\mun$ entering the magnetic energy density. At magnetic
criticality, $\mun$ becomes field dependent, which changes the thermodynamic
critical field. In addition, the leading term in the London equation vanishes,
which leads to a generalized London equation that describes exponential decay
on a length scale shorter than $\lambda_0$, in agreement with the qualitative
arguments in Sec.\ \ref{sec:II}, and with implications for the critical current
as discussed there. Unlike in the previous general discussion, in most of this
section we will treat the magnetic critical behavior in a mean-field
approximation.

\subsection{LGW theory for superconducting and magnetic fluctuations}
\label{subsec:III.A}

Our starting point is an action for a complex scalar field $\psi$ describing
the superconducting degrees of freedom coupled to a vector potential ${\bm A}$,
and a real vector field ${\bm M}$ describing the spin degrees of
freedom.\cite{Blount_Varma_1979, Radzihovsky_et_al_2001} We reiterate that the
qualitative behavior near the superconducting $T_{\text{c}}$ does not depend on
the symmetry of the order parameter, so our restriction to a scalar order
parameter does not imply a loss of generality. The action reads
\begin{widetext}
\bea
S &=& \int d{\bm x}\,\biggl[\frac{1}{2m}\,\left\vert \left({\bm\nabla} - iq{\bm
A}({\bm x})\right)\psi({\bm x})\right\vert^2 + \frac{t_1}{2}\,\vert\psi({\bm
x})\vert^2 + \frac{u_1}{4}\,\vert\psi({\bm x})\vert^4 + \frac{1}{8\pi}\,{\bm
B}^2({\bm x})
\nonumber\\
&& \hskip 30pt + \frac{a}{2}\,\left(\nabla{\bm M}({\bm x})\right)^2 +
\frac{t_2}{2}\,{\bm M}^2({\bm x}) + \frac{u_2}{4}\,\left({\bm M}^2({\bm
x})\right)^2 - {\bm M}({\bm x})\cdot{\bm B}({\bm x}) - \frac{1}{4\pi}\,{\bm
H}({\bm x})\cdot{\bm B}({\bm x})\biggr].\
\label{eq:3.1}
\eea
Here and in the remainder of this section we use units such that Planck's
constant and the speed of light are unity, $\hbar = c = 1$. The first line is
the standard Landau-Ginzburg-Wilson (LGW) functional for singlet
superconductors. The first three terms in the second line are a standard
vector-${\bm M}^4$ theory, with ${\bm M}({\bm x})$ the fluctuating
magnetization. ${\bm M}$ couples to the vector potential via the ${\bm
M}\cdot{\bm B}$ term,\cite{coupling_footnote} with ${\bm B} =
{\bm\nabla}\times{\bm A}$, and the last term is necessary to relate $S$ to the
appropriate Gibbs free energy, see Eq.\ (\ref{eq:2.2}). Notice that $\psi$ and
${\bm M}$ are coupled only indirectly via the vector potential ${\bm A}$.
Spin-flip scattering of electrons by the magnetic moments does give rise to a
direct coupling of the form ${\bm
M}^2\vert\psi\vert^2$,\cite{Greenside_Blount_Varma_1981} but these terms are
not important for our purposes.

Minimizing this action with respect to $\psi^*$, ${\bm A}$, and ${\bm M}$
yields the following saddle-point equations,
\bse
\label{eqs:3.2}
\bea
t_1\,\psi({\bm x}) + u_1\,\vert\psi({\bm x})\vert^2\psi({\bm x}) -
\frac{1}{m}\,\left({\bm\nabla} - iq{\bm A}({\bm x})\right)^2\psi({\bm x}) &=&
0,
\label{eq:3.2a}\\
-i\,\frac{q}{2m}\,\left[\psi^*({\bm x}){\bm\nabla}\psi({\bm x}) - \psi({\bm
x}){\bm\nabla}\psi^*({\bm x})\right] - \frac{q^2}{m}\,\vert\psi({\bm x})\vert^2
{\bm A}({\bm x}) &=& \frac{1}{4\pi}\,{\bm\nabla}\times\left[{\bm B}({\bm x}) -
{\bm H}({\bm x}) - 4\pi{\bm M}({\bm x})\right],
\label{eq:3.2b}\\
t_2\,{\bm M}({\bm x}) - a\,{\bm\nabla}^2{\bm M}({\bm x}) + u_2\,{\bm M}^2({\bm
x}) {\bm M}({\bm x}) &=& {\bm\nabla}\times{\bm A}({\bm x}).
\label{eq:3.2c}
\eea
\ese
\end{widetext}
If we drop Eq.\ (\ref{eq:3.2c}) and put ${\bm M}=0$ in Eq.\ (\ref{eq:3.2b})
(this corresponds to dropping ${\bm M}$ from the action) we recover the usual
Ginzburg-Landau equations.\cite{Tinkham_1975} A non-superconducting solution of
the full equations is $\psi=0$, ${\bm B} = {\bm H} + 4\pi{\bm M}$, and ${\bm
M}$ determined by the magnetic equation of state
\be
\left(r - a{\bm\nabla}^2\right){\bm M}({\bm x}) + u_2\,{\bm M}^2({\bm x}){\bm
M}({\bm x}) = {\bm H}({\bm x}),
\label{eq:3.3}
\ee
where $r = t_2 - 4\pi$. For a small constant external field ${\bm H}$ a
solution of Eq.\ (\ref{eq:3.3}) is ${\bm M} = \chin\,{\bm H}$, with
\be
\chin = 1/r
\label{eq:3.4}
\ee
the normal-state magnetic susceptibility (see Ref.\
\onlinecite{coupling_footnote}). At this point it is the bare susceptibility,
but it is clear that by renormalizing the spin part of the action before
constructing the saddle-point solution one can make it the physical
susceptibility.

\subsection{Effective theory for paramagnetic superconductors}
\label{subsec:III.B}

Now consider the full Eqs.\ (\ref{eqs:3.2}). For a small and slowly varying
${\bm M}({\bm x})$ we have from Eq.\ (\ref{eq:3.2c})
\be
{\bm M}({\bm x}) = \left(\chin^{-1} + 4\pi\right)^{-1}{\bm B}({\bm x}).
\label{eq:3.5}
\ee
Substituting this into Eq.\ (\ref{eq:3.2b}) we obtain
\bse
\label{eqs:3.6}
\be
{\bm j}_{\,\text{sc}}({\bm x}) = \frac{1}{4\pi\mun}\,{\bm\nabla}\times {\bm
B}({\bm x}) - \frac{1}{4\pi}\,{\bm\nabla}\times{\bm H}({\bm x}),
\label{eq:3.6a}
\ee
where
\bea
{\bm j}_{\,\text{sc}}({\bm x}) &=& -i\,\frac{q}{2m}\,\left[\psi^*({\bm
x}){\bm\nabla}\psi({\bm x}) - \psi({\bm x}){\bm\nabla}\psi^*({\bm x})\right]
\nonumber\\
&& - \frac{q^2}{m}\,\vert\psi({\bm x})\vert^2 {\bm A}({\bm x}).
\label{eq:3.6b}
\eea
\ese
Together with Eq.\ (\ref{eq:3.2a}), these are the equations of motion for an
effective action\cite{Chen_Lubensky_Nelson_1978}
\bea
S_{\text{eff}} &=& \int d{\bm x}\,\biggl[\frac{1}{2m}\,\left\vert
\left({\bm\nabla} - iq{\bm A}({\bm x})\right)\psi({\bm x})\right\vert^2 +
\frac{t}{2}\,\vert\psi({\bm x})\vert^2
\nonumber\\
&& + \frac{u}{4}\,\vert\psi({\bm x})\vert^4 + \frac{1}{8\pi\mun}\,{\bm
B}^2({\bm x}) - \frac{1}{4\pi}\,{\bm H}({\bm x})\cdot{\bm B}({\bm x})\biggr],
\nonumber\\
\label{eq:3.7}
\eea
where we have dropped the now-superfluous subscript on the Landau parameters
$t$ and $u$. The same result is of course obtained by starting with Eq.\
(\ref{eq:3.1}) and integrating out ${\bm M}$ in a Gaussian approximation.

The quantity ${\bm j}_{\,\text{sc}}$ in Eqs.\ (\ref{eqs:3.6}) is indeed the
supercurrent, as can be seen by comparing Eq.\ (\ref{eq:3.6a}) with Eq.\
(\ref{eq:2.9a}). It does not explicitly depend on $\mun$, see Eq.\
(\ref{eq:3.6b}), and this is important for the flux quantum to be independent
of $\mu_n$. The magnetic energy ${\bm B}^2/8\pi\mun$, which does explicitly
depend on $\mun$, does not appreciably contribute to the free energy of a thin
film or wire sample, and the standard determination of the critical current,
Ref.\ \onlinecite{Tinkham_1975}, thus leads to the usual Ginzburg-Landau result
with no correction due to $\mun\neq 1$. This corroborates the educated guess in
Sec.\ \ref{subsubsec:II.A.3}.

For all other quantities, the usual analysis of Ginzburg-Landau theory now
applies.\cite{Tinkham_1975} One characteristic length scale is given by the
square root of the ratio of the coefficients of the gradient-squared term and
the $\psi^2$ term in Eq.\ (\ref{eq:3.7}). This is the superconducting coherence
length $\xi = \sqrt{1/m\vert t\vert}$. Another one is given by the square root
of the ratio of the coefficients of the terms quadratic in ${\bm A}$. For a
constant $\psi$, this is the London penetration depth
\be
\lambda = \sqrt{m/4\pi q^2 \psi^2\mun} \equiv \lambda_0/\sqrt{\mun}.
\label{eq:3.8}
\ee
This is identical with Eq.\ (\ref{eq:2.16b}), which had been deduced on
elementary phenomenological grounds.

For the Ginzburg-Landau parameter $\kappa = \lambda/\xi$ we now have $\kappa =
\kappa_0/\sqrt{\mun}$, with $\kappa_0$ the value of the parameter for $\mun=1$.
This implies that the superconductor is of type I or type II, respectively, for
$\kappa_0 < \sqrt{\mun/2}$ or $\kappa_0 > \sqrt{\mun/2}$. While one can show
this by an explicit analysis of the effective action, a fast way to relate the
theory for arbitrary values of $\mun$ to the one for $\mun=1$ is to rewrite the
action in terms of dimensionless
quantities.\cite{Saint-James_Sarma_Thomas_1969} In conventional Ginzburg-Landau
theory, this is done by introducing
\be
{\bm x} = \lambda_0\,{\hat{\bm x}}\ ,\ \psi({\bm x}) =
\psi_0\,{\hat\psi}(\hat{\bm x})\ ,\ {\bm A}({\bm x}) =
\sqrt{2}H_{\text{c}}^0\lambda_0{\hat{\bm A}}(\hat{\bm x}),.
\label{eq:3.9}
\ee
Here $\psi_0 = \sqrt{-t/u}$ is the superconducting order parameter scale. In
terms of these quantities, the effective action
reads\cite{Saint-James_Sarma_Thomas_1969}
\bea
S_{\text{eff}} &=& \frac{(H_{\text{c}}^0)^2 \lambda_0^3}{4\pi}\int d\hat{\bm
x}\,\biggl[\left\vert \left(\frac{1}{\kappa_0}\,\hat{\bm\nabla} - i\hat{\bm
A}(\hat{\bm x})\right)\hat{\psi}(\hat{\bm x})\right\vert^2
\nonumber\\
&& - \vert\hat{\psi}(\hat{\bm x})\vert^2 +
\frac{1}{2}\,\vert\hat{\psi}(\hat{\bm x})\vert^4 +
\frac{1}{\mun}\,\left(\hat{\bm\nabla}\times\hat{\bm A}(\hat{\bm x})\right)^2
\nonumber\\
&& - 2\hat{\bm H}(\hat{\bm x})\cdot\left(\hat{\bm\nabla}\times\hat{\bm A}({\bm
x})\right),
\nonumber\\
\label{eq:3.10}
\eea
A simple further rescaling procedure shows that $S_{\text{eff}}$ depends only
on a single dimensionless parameter, rather than the two parameters $\kappa_0$
and $\mun$. Define
\be
\hat{\bm x} = \tilde{\bm x}/\kappa_0\ ,\ \hat{\bm A}(\tilde{\bm x}/\kappa_0) =
\tilde{\bm A}(\tilde{\bm x})\ ,\ \hat\psi(\tilde{\bm x}/\kappa_0) =
\tilde{\psi}(\tilde{\bm x}).
\label{eq:3.11}
\ee
Then
\bea
S_{\text{eff}} &=& \frac{(H_{\text{c}}^0)^2 \xi^3}{4\pi}\int d\tilde{\bm
x}\,\biggl[\left\vert \left(\tilde{\bm\nabla} - i\tilde{\bm A}(\tilde{\bm
x})\right)\tilde{\psi}(\tilde{\bm x})\right\vert^2
\nonumber\\
&& - \vert\tilde{\psi}(\tilde{\bm x})\vert^2 +
\frac{1}{2}\,\vert\tilde{\psi}(\tilde{\bm x})\vert^4 +
\frac{\kappa_0^2}{\mun}\,\left(\tilde{\bm\nabla}\times\tilde{\bm A}(\tilde{\bm
x})\right)^2
\nonumber\\
&& - 2\frac{\kappa_0}{\sqrt{\mun}}\,\left(\sqrt{\mun}\tilde{\bm H}(\tilde{\bm
x})\right)\cdot\left(\tilde{\bm\nabla}\times\tilde{\bm A}(\tilde{\bm
x})\right).
\nonumber\\
\label{eq:3.12}
\eea
This shows that the theory with an arbitrary $\mun$ maps onto ordinary
Ginzburg-Landau theory with the replacements
\be
\kappa_0 \to \kappa_0/\sqrt{\mun} \equiv \kappa\quad,\quad {\bm H} \to
\sqrt{\mun}\,{\bm H}.
\label{eq:3.13}
\ee
$\kappa_0 = \sqrt{\mun/2}$ thus indeed marks the demarcation between type I and
type II superconductors, and the critical fields can be immediately obtained
from the usual results at $\mun=1$.\cite{Tinkham_1975} For the thermodynamic
critical field $H_{\text{c}}$, the upper critical field $H_{\text{c}2}$, the
lower critical field $H_{\text{c}1}$, and the surface critical field
$H_{\text{c3}}$ we obtain
\bse
\label{eqs:3.14}
\bea
H_{\text{c}} &=& H_{\text{c}}^0/\sqrt{\mun},
\label{eq:3.14a}\\
H_{\text{c}2} &=& H_{\text{c}2}^0/\mun = \sqrt{2} \kappa_0 H_{\text{c}}^0/\mun,
\label{eq:3.14b}\\
H_{\text{c}1} &=& H_{\text{c}1}^0\,\frac{g(\kappa_0/\sqrt{\mun})}{g(\kappa_0)}
= \frac{H_{\text{c}}^0}{\sqrt{2}\kappa_0}\,g(\kappa_0/\sqrt{\mun}).\qquad
\label{eq:3.14c}\\
H_{\text{c}3} &=& 1.695\,H_{\text{c}2}.
\label{eq:3.14d}
\eea
where the universal function $g$ has the limiting behavior
\be
g(x) = \begin{cases} \ln x + 0.08 + O(1/x) & \text{for}\quad x\gg 1/\sqrt{2},
                                                                           \cr
                     1                     & \text{for}\quad x = 1/\sqrt{2}.
       \end{cases}
\label{eq:3.14e}
\ee
\ese
If one neglects the weak dependence of $g$ on its argument, $H_{\text{c}1}$ is
approximately independent of $\mun$.

\subsection{Superconductors at magnetic criticality}
\label{subsec:III.C}

As one approaches a ferromagnetic instability, $\mun$ keeps increasing and can
no longer be treated as a constant. There are two effects that become important
for our purposes. First, in a normal metal $\mun$ becomes strongly field or
induction dependent. At $r=0$ this dependence is nonanalytic and described by
the critical exponent $\delta$. Second, as $r$ becomes on the order of
$\xi_{\text{m}}^0/\lambda$ (see Sec.\ \ref{subsec:II.B}) in a superconducting
phase, the difference between $\mun$ and $\mus$ can no longer be neglected.
Related to this, the gradient squared term in Eq.\ (\ref{eq:3.2c}) must be
taken into account. We now consider these effects, starting with the
nonanalytic field dependence in the normal state.

\subsubsection{Thermodynamic critical field}
\label{subsubsec:III.C.1}

At magnetic criticality in the normal state, $r=0$, one has\cite{Ma_1976}
\be
\chin (r=0,H) = \chi_0\left(H/{\tilde H}_0\right)^{1/\delta - 1}.
\label{eq:3.15}
\ee
Here $\chi_0$ is a microscopic susceptibility, and ${\tilde H}_0$ is a
microscopic field scale. $M$ and, for small values of $H$, $B$ are therefore
proportional to $H^{1/\delta}$, or $H\propto B^{\delta}$. For small $B$, the
number $\mun$ should thus be replaced by a function of $B$ with the following
leading $B$-dependence,
\be
\mun \to (H_0/B)^{\delta - 1},
\label{eq:3.16}
\ee
with $H_0 = (4\pi\chi_0)^{\delta/(\delta - 1)} \tilde{H}_0$. The magnetic
energy cost of the flux expulsion that results from the formation of a Meissner
phase (which equals minus the normal-state magnetic energy) is now obtained by
using Eq.\ (\ref{eq:3.16}) in Eq.\ (\ref{eq:3.7}). It is $E_{\text{m}}/V =
HB/4\pi - B^{\delta+1}/8\pi H_0^{\delta-1} = H_0^{1-1/\delta}
H^{1+1/\delta}/8\pi$. The condensation energy is still given by
$E_{\text{cond}}/V = t^2/4u$, which yields
\be
\Hc = \left(\frac{2\pi}{u}\right)^{\delta/(\delta+1)}
\frac{1}{H_0^{(\delta-1)/(\delta+1)}}\ \vert t\vert^{2\delta/(\delta+1)}.
\label{eq:3.17}
\ee
The thermodynamical critical field is thus weaker than in the paramagnetic
case, and the $t$-dependence is consistent with Eq.\ (\ref{eq:2.21}). By
comparing with Eq.\ (\ref{eq:2.5}), we see that with respect to the
thermodynamical critical field, $\mun$ effectively scales like $\mun \sim
1/\vert t\vert^{2(\delta-1)/(\delta+1)}$ at magnetic criticality.

Equations (\ref{eq:3.15}) through (\ref{eq:3.17}) hold also for small but
nonzero values of $r$ as long as one is in the field scaling regime, i.e, as
long as $H$ in appropriate units is large compared to $r$ to an appropriate
power. We will discuss this in more detail in Sec.\ \ref{sec:IV}. At this point
we only mention that, since $\Hc$ vanishes as $\vert t\vert \to 0$,
sufficiently close to $\Tc$ one will lose the field scaling for any nonzero
value of $r$, and $\Hc$ will be given by Eq.\ (\ref{eq:3.14a}).

\subsubsection{Generalized London equation}
\label{subsubsec:III.C.2}

The ordinary London equation is obtained from Eq.\ (\ref{eq:3.2b}) by dropping
${\bm M}({\bm x})$ and treating $\psi({\bm x}) \equiv \psi$ as a constant
(London approximation). With ${\bm\nabla}\times{\bm H}({\bm x}) = 0$ this leads
to
\be
-\lambda_0^{-2}\,{\bm B}({\bm x}) = {\bm\nabla}\times{\bm\nabla}\times{\bm
B}({\bm x}).
\label{eq:3.18}
\ee
Now take ${\bm M}$ into account. Using Eq.\ (\ref{eq:3.2c}) in Eq.\
(\ref{eq:3.2b}), we can eliminate ${\bm B}$ and derive an equation for ${\bm
M}$. Once ${\bm M}$ is known, ${\bm B}$ follows from Eq.\ (\ref{eq:3.2c}).
Within the London approximation one finds
\begin{widetext}
\bea
{\bm M}({\bm x}) &=& -(\lambda_0^2/\mun){\bm\nabla}\times{\bm\nabla}\times{\bm
M}({\bm x}) + \left({\tilde\xi}_{\text{m}}^0\right)^2{\bm\nabla}^2{\bm M}({\bm
x}) +
\left({\tilde\xi}_{\text{m}}^0\right)^2\lambda_0^2{\bm\nabla}\times{\bm\nabla}\times
 {\bm\nabla}^2{\bm M}({\bm x}) - {\tilde u}\,{\bm M}^2({\bm x}){\bm M}({\bm x})
 \nonumber\\
 && - {\tilde u}\lambda_0^2{\bm\nabla}\times{\bm\nabla}\times{\bm M}^2({\bm x}){\bm
       M}({\bm x}).
\label{eq:3.19}
\eea
\end{widetext}
Here $\mun = (4\pi + r)/r$ as in Sec.\ \ref{subsec:III.B},
${\tilde\xi}_{\text{m}}^0 = \xi_{\text{m}}^0/\sqrt{4\pi + r} \equiv
\sqrt{a/(4\pi + r)}$, and ${\tilde u} = u/(4\pi + r)$.

As long as $\mun \approx 1$, the first term on the right-hand side of Eq.\
(\ref{eq:3.19}) leads to a variation of ${\bm M}$ on a length scale $\lambda =
\lambda_0/\sqrt{\mun}$. The second term is a small correction to the first one
since $\xi_{\text{m}}^0 \ll \lambda_0$. So is the third term, which is of order
$\left({\tilde\xi}_{\text{m}}^0\right)^2{\bm\nabla}^2 \sim
\left({\tilde\xi}_{\text{m}}^0\right)^2/\lambda_0^2 \ll 1$ relative to the
first one. The linearized version of Eq.\ (\ref{eq:3.19}) thus reduces to the
ordinary London equation, Eq.\ (\ref{eq:3.18}), with $\lambda_0 \to \lambda$.
However, for $r=0$ the first term vanishes. This makes the second term the
leading one, and the third term, which is of order $\lambda_0^2{\bm\nabla}^2$
compared to the second one, cannot be neglected either. The linearized equation
thus reads
\be
{\bm M}({\bm x}) = \left({\tilde\xi}_{\text{m}}^0\right)^2{\bm\nabla}^2 \left[1
+ \lambda_0^2{\bm\nabla}\times{\bm\nabla}\times\right]{\bm M}({\bm x}).
\label{eq:3.20}
\ee
With the same interface geometry as in Sec.\ \ref{subsubsec:II.A.2} this takes
the form
\be
M(x) = \left({\tilde\xi}_{\text{m}}^0\right)^2 M''(x) -
\left({\tilde\xi}_{\text{m}}^0\right)^2 \lambda_0^2 M^{(iv)}(x).
\label{eq:3.21}
\ee
This linear quartic ODE is solved by an exponential ansatz, $M(x) = M_0
e^{-\rho x}$. The real solution that falls off for $x\to\infty$ shows damped
oscillatory behavior. From Eq.\ (\ref{eq:3.2c}) we see that $B(x)$ shows the
same behavior as $M(x)$, up to corrections of
$O({\tilde\xi}_{\text{m}}^0/\lambda_0)$. With the boundary condition $B(x=0) =
\mun H$ we finally obtain
\be
B(x) = \mun H e^{-x/\sqrt{2{\tilde\xi}_{\text{m}}^0\lambda_0}} \cos
\left(x/\sqrt{2{\tilde\xi}_{\text{m}}^0\lambda_0}\ \right)\ .
\label{eq:3.22}
\ee

This is the solution of the linearized version of Eq.\ (\ref{eq:3.19}) at
$r=0$. In addition to leaving out the terms of $O(M^3)$, we have also ignored
the fact that the permeability, whether $\mun$ or $\mus$, does depend on $B$ or
$M$ at magnetic criticality. In a mean-field approximation, $\mun \propto
1/B^2$ at $r=0$, see Eq.\ (\ref{eq:3.16}), which also leads to terms of
$O(M^3)$ in the nonlinear equation. Depending on the ratio of the external
field to $H_0$, these terms may or may not be important for the initial decay
of $M$ or $B$ near the normal metal-to-superconductor boundary. However, once
$M$ or $B$ has decayed sufficiently, these terms become subleading compared to
the linear ones in Eq.\ (\ref{eq:3.21}), and the asymptotic behavior as $B\to
0$ is always given by Eq.\ (\ref{eq:3.22}).

In order to make contact with the discussion in Sec.\ \ref{subsec:II.B} for
small but nonzero values of $r$, let us consider the linearized Eq.\
(\ref{eq:3.19}) while keeping the first term. Instead of Eq.\ (\ref{eq:3.21})
we then have
\be
M(x) = \left(\lambda_0^2/\mun + \left({\tilde\xi}_{\text{m}}^0\right)^2\right)
M''(x) - \left({\tilde\xi}_{\text{m}}^0\right)^2 \lambda_0^2 M^{(iv)}(x).
\label{eq:3.23}
\ee
This is solved by
\bse
\label{eqs:3.24}
\be
M(x) = M_0\,e^{\rho x},
\label{eq:3.24a}
\ee
with
\bea
\rho^2 &=& \frac{1}{2\lambda_0^2 \left({\tilde\xi}_{\text{m}}^0\right)^2}
 \biggl[\lambda_0^2/\mun + \left({\tilde\xi}_{\text{m}}^0\right)^2\hskip 90pt
 \nonumber\\
 && \hskip 0pt - \sqrt{\left(\lambda_0^2/\mun +
 \left({\tilde\xi}_{\text{m}}^0\right)^2\right)^2 - 4\lambda_0^2
 \left({\tilde\xi}_{\text{m}}^0\right)^2}\biggr].
\label{eq:3.24b}
\eea
\ese
Here we have chosen the solution for $\rho^2$ that yields $\rho^2 \to
1/\lambda_0^2$ for $r\to\infty$. Equation\ (\ref{eq:3.24b}) still provides two
solutions for $\rho$, and the physical solution for $M$ is determined by the
requirement that $M$ be real.

A discussion of Eq.\ (\ref{eq:3.24b}) shows that $\rho^2$ becomes purely real
and negative at $r = r_{\text{s}} = -4\sqrt{\pi}
{\tilde\xi}_{\text{m}}^0/\lambda_0 +
O\left(\left({\tilde\xi}_{\text{m}}^0\right)^2/\lambda_0^2\right)$. This is in
agreement with the results of Blount and Varma,\cite{Blount_Varma_1979} who
showed that spiral magnetic order coexisting with the superconductivity occurs
at this point. For $\vert r\vert \ll {\tilde\xi}_{\text{m}}^0/\lambda_0$ one
has $\rho^2 \approx -i/{\tilde\xi}_{\text{m}}^0 \lambda_0$, which leads to Eq.\
(\ref{eq:3.22}). For $r \gg {\tilde\xi}_{\text{m}}^0/\lambda_0$ one finds
$\rho^2 \approx \mun/\lambda_0^2$, which leads to
\be
B(x) = \mun H e^{-x\sqrt{\mun}/\lambda_0},
\label{eq:3.25}
\ee
in agreement with Eq.\ (\ref{eq:3.8}).

\subsubsection{Penetration depth, and critical current}
\label{subsubsec:III.C.3}

Equation (\ref{eq:3.22}) shows that the effective penetration depth at magnetic
criticality is
\be
\lambda = \sqrt{2{\tilde\xi}_{\text{m}}^0\lambda_0}\quad,\quad (\vert r\vert
\ll {\tilde\xi}_{\text{m}}^0/\lambda_0),
\label{eq:3.26}
\ee
in agreement with the conclusions of Ref.\
\onlinecite{Greenside_Blount_Varma_1981} drawn from studying the ferromagnetic
phase, and with Eq.\ (\ref{eq:2.22}) with $\gamma = 1$. The latter
approximation results from the fact that our saddle-point equations of motion
describe the magnetic equation of state in a mean-field approximation. The
discussion of Eq. (\ref{eq:3.24b}) shows that this result is valid for $\vert
r\vert \ll {\tilde\xi}_{\text{m}}^0/\lambda_0$. By comparing with Eq.\
(\ref{eq:2.16b}) or (\ref{eq:3.8}), we see that with respect to the penetration
depth, $\mun$ at magnetic criticality scales like $\mun \sim 1/\sqrt{\vert
t\vert}$ in mean-field approximation, or $\mun \sim 1/\vert t\vert^{\gamma/2}$
in general. The fact that $1/\sqrt{\mun}$ in Eqs.\ (\ref{eq:2.5}) and
(\ref{eq:2.16b}), respectively, must be interpreted differently for $\mun \to
\infty$ is a consequence of the influence of the superconductivity on the spin
response.

For $r \gg {\tilde\xi}_{\text{m}}^0/\lambda_0$ we have, from Eq.\
(\ref{eq:3.25})
\be
\lambda = \lambda_0/\sqrt{\mun},
\label{eq:3.27}
\ee
in agreement with Eq.\ (\ref{eq:3.8}).

The expression for the critical current given by Eq.\ (\ref{eq:2.18}) is
general within the London approximation. We have now given a derivation of the
behavior of the thermodynamical critical field and the penetration depth given
on phenomenological grounds in Eqs.\ (\ref{eq:2.21}) and (\ref{eq:2.22}),
respectively. The behavior of the critical current at or near magnetic
criticality is thus given by Eqs.\ (\ref{eqs:2.23}).

\subsubsection{Critical field $H_{\text{c2}}$}
\label{subsubsec:III.C.4}

The critical exponent $\gamma$ is positive ($\gamma \approx 1.4$ for typical
ferromagnetic universality classes in three dimensions\cite{Zinn-Justin_1996}).
The result for $\lambda$, Eq.\ (\ref{eq:2.22}) or (\ref{eq:3.23}) in mean-field
approximation, of the previous subsection therefore means that $\lambda$
diverges more slowly for $\vert t\vert \to 0$ than the superconducting
coherence length $\xi \propto 1/\sqrt{\vert t\vert}$. Consequently,
superconductors at magnetic criticality ($\vert r\vert \ll
{\tilde\xi}_{\text{m}}^0/\lambda_0$) are necessarily of type
I.\cite{Greenside_Blount_Varma_1981}

This observation notwithstanding, the critical field $H_{\text{c2}}$, which in
a type-II superconductor signalizes the boundary of the vortex phase, still has
a physical meaning: It is the minimum field to which the normal metal can be
`supercooled' before it discontinuously develops a nonzero superconducting
order parameter.\cite{Tinkham_1975} It is thus still of interest to determine
$H_{\text{c2}}$. Furthermore, the behavior will be necessarily of type I only
for $\vert r\vert$ in an extremely narrow region. Outside of this region, Eq.\
(\ref{eq:3.27}) holds, and for a sufficiently large value of $\kappa_0 =
\lambda_0/\xi$ the superconductor will still be of type II. The determination
of $H_{\text{c2}}$ is done by linearizing the Ginzburg-Landau equation, Eq.\
(\ref{eq:3.2a}), in $\psi$. It then turns into a Schr{\"o}dinger equation for a
particle in a vector potential ${\bm A}$, with $-t_1/2 \equiv -t/2$ playing the
role of the energy eigenvalue. By means of standard
arguments\cite{Tinkham_1975} this leads to a critical value of the magnetic
induction ${\bm B} = {\bm\nabla}\times{\bm H}$ given by $B_{\text{c2}} \equiv
H_{\text{c2}}^0 = -tm/q$. In a paramagnetic superconductor, this leads to
\be
H_{\text{c2}} = H_{\text{c2}}^0/\mun \quad (\mun = \text{const.}),
\label{eq:3.28}
\ee
which is the same as Eq.\ (\ref{eq:3.14b}). At magnetic criticality, we have,
cf. Eq.\ (\ref{eq:3.16}),
\be
H_{\text{c2}} = B_{\text{c2}}^{\delta}/H_0^{(\delta - 1)} \propto \vert
t\vert^{\delta}.
\label{eq:3.29}
\ee
Notice that, in this context, $\mun$ scales as $\mun \sim 1/\vert
t\vert^{\delta - 1}$, whereas it scales as $\mun \sim 1/\vert t\vert$ if the
relevant field scale is $H_{\text{c}}$. Since $H_{\text{c2}}$ vanishes much
faster than $\Hc$, Eq.\ (\ref{eq:3.17}), the field scaling region will be
restricted to larger values of $\vert t\vert$, and $H_{\text{c2}}$ will be
given by Eq.\ (\ref{eq:3.28}) in a substantial range of $t$-values. Will come
back to this in Sec.\ \ref{sec:IV}.

\section{Discussion and Conclusion}
\label{sec:IV}

To summarize, we have determined the electrodynamic properties of
superconductors close to a ferromagnetic instability, i.e., materials that, in
the absence of superconductivity, would be paramagnetic with large
ferromagnetic fluctuations. This work complements previous studies of the
coexistence of superconductivity with ferromagnetic
order.\cite{Blount_Varma_1979, Greenside_Blount_Varma_1981} We have treated the
superconductivity in mean-field (Ginzburg-Landau) approximation. In addition,
we have employed the London approximation, treating the superconducting order
parameter as a constant. The ferromagnetic critical point we have treated
explicitly in a mean-field approximation, and we have used scaling arguments to
consider the consequences of the exact magnetic critical behavior for the
superconductivity. We have found that the thermodynamical critical field $\Hc$
decreases due to the ferromagnetic fluctuations, as one would expect, and
depends on the magnetic critical exponent $\delta$, see Eqs.\ (\ref{eq:3.17})
and (\ref{eq:2.21}). However, the London penetration depth also decreases,
which is intuitively less obvious. At magnetic criticality the behavior of the
magnetic induction at a vacuum-to-superconductor (or normal
metal-to-superconductor) interface is still characterized by exponential decay,
but the characteristic length scale $\lambda$ is different from the usual
London penetration depth $\lambda_0$. Within a mean-field description of the
magnetic criticality it is the geometric mean of the zero-temperature magnetic
correlation length and $\lambda_0$, see Eqs.\ (\ref{eq:3.22}) and
(\ref{eq:3.26}); more generally, it depends on the magnetic critical exponent
$\gamma$, see Eq.\ (\ref{eq:2.22}). However, this behavior of the penetration
depth is valid only within an extremely small region of width
${\tilde\xi}_{\text{m}}^0/\lambda_0$ around magnetic criticality. Outside of
this region, but still within the ferromagnetic critical region, the
temperature dependence of the penetration depth is the same as in
Ginzburg-Landau theory, see Eq.\ (\ref{eq:3.27}). For the critical current $\jc
\propto \Hc/\lambda$ this implies a dependence on the reduced temperature given
by $\vert t\vert^{\alpha}$, where the exponent $\alpha$ depends on both
$\delta$ and $\gamma$, or on $\delta$ only, depeding on the value of $r$, see
Eqs.\ (\ref{eqs:2.23}). With exponent values appropriate for the usual
ferromagnetic universality classes, $\alpha \approx 1.8$ extremely close to
magnetic criticality, and $\alpha \approx 2.15$ somewhat farther away.

Let us now discuss these results in some more detail, and relate them to the
experimental observations reported in Ref.\
\onlinecite{Young_Moldovan_Adams_2004}.

\begin{figure}[t,h]
\vskip -0mm
\includegraphics[width=7.0cm]{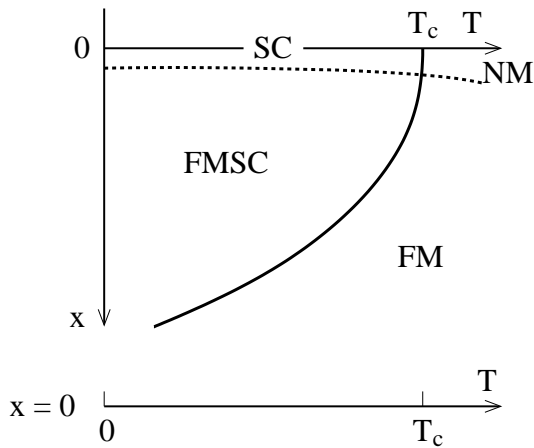}
\caption{Schematic phase diagram showing a normal metal (NM), a ferromagnet
 (FM), a superconductor (SC), and a ferromagnetic superconductor (FMSC) in a
 temperature ($T$) - control parameter ($x$ plane. The solid line denotes the
 superconducting transition, the dashed line, the magnetic one. Along $x=0$
 there is only one phase transition at the superconducting $\Tc$. See the text
 for additional explanation.}
\label{fig:2}
\end{figure}

\begin{figure}[t]
\vskip -0mm
\includegraphics[width=7.0cm]{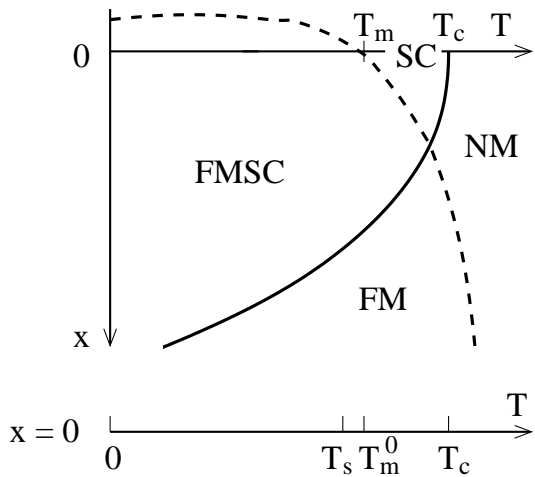}
\caption{Same as Fig.\ \ref{fig:2}, but with a magnetic transition for $x=0$ at
a temperature $T_{\text{m}} < \Tc$. On the $x=0$ axis it is shown that
$T_{\text{m}}$ splits into the bare magnetic transition temperature
$T_{\text{m}}^0$ and the physical transition temperature $T_{\text{s}}$ to a
state with spiral magnetic order, Ref.\onlinecite{Blount_Varma_1979}. See the
text for additional explanation.}
\label{fig:3}
\end{figure}

For the temperature dependencies of various observables at magnetic criticality
we have assumed that the system stays tuned to magnetic criticality while the
temperature is varied. Let us discuss to what extent this assumption is
realistic. Consider a phase diagram in a plane spanned by the temperature and
some non-thermal control parameter $x$, e.g., the hole doping concentration in
the case of MgCNi$_3$,\cite{Rosner_et_al_2002} and consider the following two
qualitatively different possibilities. Figure \ref{fig:2} shows a situation
where the magnetic phase separation line does not cross the line $x=0$. The
stoichiometric compound thus does not enter a magnetic phase upon cooling,
although the system is close to a magnetic transition for all temperatures
below the superconducting $\Tc$. This scenario is believed to apply to
MgCNi$_3$.
Figure \ref{fig:3} shows a situation where the magnetic phase separation line
does cross the line $x=0$, so that the stoichiometric compound enters a phase
where superconductivity and magnetism coexist at some temperature below $\Tc$.
This is the situation that was discussed in Refs.\
\onlinecite{Blount_Varma_1979} and \onlinecite{Greenside_Blount_Varma_1981} and
observed in ErRh$_4$B$_4$ and HoMo$_6$S$_8$.\cite{Moncton_et_al_1980,
Lynn_et_al_1981}
The magnetic transition is to a phase with spiral magnetic order at a
temperature $T_{\text{s}}$ slightly below the temperature $T_{\text{m}}^0$
where ferromagnetism would occur in the absence of
superconductivity.\cite{Blount_Varma_1979}

We now can see what is required to keep $r$ constant while varying $t$, namely,
a situation as shown in Fig.\ \ref{fig:2} with the dashed line essentially
parallel to the $T$-axis. $r$ is then given by the dimensionless distance
between the two lines. In order for the penetration depth to display the
non-Ginzburg-Landau behavior described by Eq.\ (\ref{eq:3.26}) or, more
generally, Eq.\ (\ref{eq:2.22}), the two lines would have to be extremely
close, in order to keep $r$ smaller than ${\tilde\xi}_{\text{m}}^0/\lambda_0$,
see Eq.\ (\ref{eq:3.26}). This would result in a temperature dependence of the
critical current given by Eqs.\ (\ref{eq:2.23a}, \ref{eq:2.23b}). While this is
possible, it is a very non-generic situation, and it would result in a very
large magnetic susceptibility of the normal metal just above the
superconducting transition temperature.

A situation that is still very non-generic, but requires somewhat less
fine-tuning, is one where the dashed line is still essentially parallel to the
$T$-axis, but in a somewhat larger $r$-range, say, with $r$ on the order of a
few percent. In this case the penetration depth will show the usual $1/\vert
t\vert^{1/2}$ temperature dependence, see Eq.\ (\ref{eq:3.27}). The temperature
dependence of the thermodynamic critical field will be more complicated in this
case. The generalization of Eq.\ (\ref{eq:2.20}) to nonzero values of $r$ is
\be
\chin = r^{-\gamma}\,f_{\chi}(H/r^{\beta\delta}),
\label{eq:4.1}
\ee
with $\gamma = \beta (\delta - 1)$, $\beta$, and $\delta$ the usual critical
exponents for the magnetic transition. In order for Eq.\ (\ref{eq:2.20}) to
hold, the $H$ must be large compared to $r^{\beta\delta}$ in suitable units.
The latter are not determined by any universal arguments, but an analysis of
the critical equation of state for both the high-temperature ferromagnet Ni
($T_{\text{m}} \approx 630\,{\text{K}}$)\cite{Kouvel_Comly_1968} and the
low-temperature ferromagnet CrBr$_3$ ($T_{\text{m}} \approx
33\,{\text{K}}$)\cite{Ho_Litster_1969} shows that in either case the relevant
energy or field scale (we use units such that $k_{\text{B}} = \mu_{\text{B}} =
1$) is given by $T_{\text{m}}$, which is plausible. The crossover between the
field scaling that leads to Eq.\ (\ref{eq:2.21}) and the static scaling that
leads to Eq.\ (\ref{eq:2.5}) thus occurs at a crossover field
\be
H_{\times} \approx T_{\text{m}}^0\,r^{\beta\delta}.
\label{eq:4.2}
\ee
$\beta\delta \approx 5/3$ for ferromagnetic phase transitions, and with
$T_{\text{m}}\approx 10\,\text{K}$ and $r\approx 0.1$, one finds $H_{\times}
\approx 0.02\,T_{\text{m}}^0$. For MgCNi$_3$ in the vicinity of $\Tc$, this
leads to $H_{\times} \approx 0.2\,\text{T}$. With $H_{c2}$ at {\em zero
temperature} on the order of $14\,\text{T}$ and $\kappa \approx
40$,\cite{Young_et_al_2003} one expects $\Hc(T=0) = H_{c2}/\sqrt{2}\kappa
\approx 0.25\,\text{T}$. Since $\Hc$ vanishes at $\Tc$, this means that $\Hc$
will be given by Eq.\ (\ref{eq:2.21}) sufficiently far away from $\Tc$, but
cross over to $\Hc \propto \vert t\vert$ near $\Tc$. Consequently, the critical
current exponent $\alpha$ will be given by Eq.\ (\ref{eq:2.23c}) at some
distance from $\Tc$, and cross over to the Ginzburg-Landau result $\alpha =
3/2$ as $\vert t\vert \to 0$. In the experiment of Ref.\
\onlinecite{Young_Moldovan_Adams_2004}, no such crossover was observed down to
$\vert t\vert \approx 0.01$.

At least within the London approximation, our results confirm the conclusion of
Ref.\ \onlinecite{Greenside_Blount_Varma_1981} that superconductors near a
ferromagnetic instability are necessarily of type I. However, we have also
shown that this conclusion is inevitable only within an extremely small region
around the (bare) magnetic critical point. The fact that MgCNi$_3$ is observed
to be of type II\cite{Young_et_al_2003} is therefore not necessarily in
contradiction to the notion that this material is almost ferromagnetic.
However, Eq.\ (\ref{eq:3.29}) predicts a strong deviation from Ginzburg-Landau
behavior for the upper critical field $H_{\text{c2}}$. Since $H_{\text{c2}}$
goes to zero rapidly as $\vert t\vert \to 0$, this behavior will show only at
substantial values of $\vert t\vert$ even if $r$ is very small. No anomalous
behavior was observed for $\vert t\vert$ up to $0.5$.\cite{Young_et_al_2003}
This is reconcilable with close proximity to a magnetic instability only if $r$
is very small close to $\Tc$, and grows with decreasing temperature, in which
case $H_{\text{c2}}$ might never show the magnetic critical behavior. A
signature of this situation would be a large magnetic susceptibility in the
normal state just above $\Tc$.

The conclusion from this discussion with respect to the experimental
observations in Ref.\ \onlinecite{Young_Moldovan_Adams_2004} is as follows.
While it is possible that proximity to a ferromagnetic instability is the cause
of the observed anomalous behavior of the critical current, such an explanation
requires fine tuning of the phase diagram, and would have to be accompanied by
a very large enhancement of the spin susceptibility in the normal phase just
above $\Tc$. Explaining the lack of an anomaly in the temperature dependence of
$H_{\text{c2}}$ probably requires that the material is closer to the magnetic
instability near $\Tc$ than at $T=0$ (i.e., the dashed line in Fig.\
\ref{fig:2} comes closer to the $T$-axis with increasing $T$). A direct
measurement of the spin susceptibility in the normal phase would be of great
interest in this context.

Finally, we discuss our predictions for the case of a superconductor that does
undergo a transition to a magnetic state below $\Tc$, i.e., the situation
represented by Fig.\ \ref{fig:3}. In the (very small) temperature interval of
width $2\vert T_{\text{m}}^0 - T_{\text{s}}\vert$ around $T_{\text{m}}^0$, both
the thermodynamic critical field $\Hc$ and the penetration depth $\lambda$ will
show an anomalous temperature dependence, and the critical current exponent
will be given by Eq.\ (\ref{eq:2.23b}). Outside of this region, but not too
close to $\Tc$, $\Hc$ will be anomalous, but $\lambda$ will be conventional,
and the critical current exponent will be given by Eq.\ (\ref{eq:2.23c}). Upon
approaching $\Tc$, $\Hc$ will fall below the crossover field given by Eq.\
(\ref{eq:4.2}), and its temperature dependence will cross over to the usual
linear Ginzburg-Landau behavior. The critical current exponent close to $\Tc$
will thus be the conventional $\alpha = 3/2$. The location of this crossover
depends on the critical field scale, and will thus be material dependent.
Critical current measurements in the materials like ErRh$_4$B$_4$, or
HoMo$_6$S$_8$, which are believed to fall into this class, would be very
interesting.

\acknowledgments We thank Phil Adams for alerting us to his results on
MgCNi$_3$, for sharing his data prior to publication, and for helpful
correspondence. This work was supported by the NSF under grant Nos.
DMR-05-29966, and DMR-05-30314.


\begin{thebibliography}{28}
\bibitem[{cou()}]{coupling_footnote}
\bibinfo{note}{One might consider it more physical to write ${\bm M}\cdot({\bm
  B} - 4\pi{\bm M})$, which would make $t_2$ the inverse (normal) magnetic
  susceptibility. However, this just amounts to a shift of $t_2$ by $4\pi$.}
\end{thebibliography}

\end{document}